\documentclass[10pt,a4paper,onecolumn]{article}
\usepackage{pdfpages}           
\usepackage{booktabs}
\PassOptionsToPackage{hyphens}{url}\usepackage{hyperref}
\usepackage{upgreek}
\usepackage{authblk}
\usepackage{float}
\usepackage{tabu}
\usepackage{amsmath}
\usepackage{rotating}
\newcommand*\samethanks[1][\value{footnote}]{\footnotemark[#1]}
\usepackage{graphicx}
\usepackage{xcolor}
\usepackage{gensymb}
\usepackage[margin=0.7in]{geometry}
\usepackage{subcaption}
\usepackage{mathtools}
\usepackage{afterpage}
\usepackage[labelformat=simple]{subcaption}

\usepackage{titlesec}

\begin{document}
\title{Observations of Pressure Anisotropy Effects within Semi-Collisional Magnetized-Plasma Bubbles}
\author[1,2,3]{E. R. Tubman \thanks{These authors contributed equally.}}
\author[4,5,6]{A. S. Joglekar \samethanks}
\author[7]{A. F. A. Bott}
\author[8]{M. Borghesi}
\author[8]{B. Coleman}
\author[9]{G. Cooper}
\author[9,3,7]{C. N. Danson}
\author[1]{P. Durey}
\author[9]{J. M. Foster}
\author[9]{P. Graham}
\author[7]{G. Gregori}
\author[9,3]{E. T. Gumbrell}
\author[9]{M. P. Hill}
\author[8]{T. Hodge}
\author[8]{S. Kar}
\author[3]{R. J. Kingham}
\author[1,10]{M. Read}
\author[1]{C. P. Ridgers}
\author[9,10]{J. Skidmore}
\author[11]{C. Spindloe}
\author[6]{A. G. R. Thomas}
\author[9]{P. Treadwell}
\author[1]{S. Wilson}
\author[6]{L. Willingale}
\author[1]{N. C. Woolsey}
\affil[1]{York Plasma Institute,  Department of Physics, University of York, UK}
\affil[2]{Lawrence Livermore National Laboratory, USA}
\affil[3]{Department of Physics, Imperial College London, UK}
\affil[4]{University of California, Los Angeles, Los Angeles, CA, USA}
\affil[5]{Noble.AI, San Francisco, USA}
\affil[6]{Gérard Mourou Center for Ultrafast Optical Science, University of Michigan, Ann Arbor, MI, USA}
\affil[7]{Department of Physics, University of Oxford, UK}
\affil[8]{School of Mathematics and Physics, Queens University Belfast, UK}
\affil[9]{AWE Aldermaston, Reading, UK}
\affil[10]{First Light Fusion, Oxford, UK}
\affil[11]{Target Fabrication, Central Laser Facility, Rutherford Appleton Laboratory, UK}

\maketitle

\abstract{\textbf{Magnetized plasma interactions are ubiquitous in astrophysical and laboratory plasmas. Various physical effects have been shown to be important within colliding plasma flows influenced by opposing magnetic fields, however, experimental verification of the mechanisms within the interaction region has remained elusive. Here we discuss a laser-plasma experiment whereby experimental results verify that Biermann battery generated magnetic fields are advected by Nernst flows and anisotropic pressure effects dominate these flows in a reconnection region. These fields are mapped using time-resolved proton probing in multiple directions. Various experimental, modelling and analytical techniques demonstrate the importance of anisotropic pressure in semi-collisional, high-$\beta$ plasmas, causing a reduction in the magnitude of the reconnecting fields when compared to resistive processes. Anisotropic pressure dynamics are crucial in collisionless plasmas, but are often neglected in collisional plasmas.  We show pressure anisotropy to be essential in maintaining the interaction layer, redistributing magnetic fields even for semi-collisional, high energy density physics (HEDP) regimes.}}\\

\section*{Introduction}
Magnetic fields of $10^{-4}-100$ T ($10-10^6$ G) can be embedded in both astrophysical plasmas and laboratory produced plasmas. An important phenomenon observed in these plasmas is magnetic reconnection, whereby magnetic fields are rapidly reorganised when plasma flows containing opposing magnetic fields are driven together. These scenarios where magnetic reconnection may occur are present in many environments ranging from the low $\beta$ plasmas where magnetic pressure dominates such as in solar flares \cite{Masuda1994}, to higher $\beta$ plasmas in the magnetospheres of planets \cite{Cassak2007,Masters2012} and low luminosity accretion flows \cite{Ball2017}. Higher $\beta$ plasmas are accessible within the laboratory using high-power lasers. The mechanisms via which the magnetic fields interact in these scenarios depends on the plasma parameters. In the experiments presented here, we detail these mechanisms for higher $\beta$ plasmas.

In this article we present proton deflectometry data, supported by simulations and detailed theory, from a study designed to observe magnetic reconnection within a high-$\beta$ ($\beta \sim 1 - 100$) plasma. Reconnection geometries of high-$\beta$ plasmas have previously been studied using lasers \cite{Rosenberg2015,Li2007,Nilson2006} where the plasma conditions are close to collisionless, i.e. $L/\lambda_\text{mfp}\leq1$. The past experimental data \cite{Rosenberg2015,Nilson2006} from these investigations shows streaks in the shadowgraphy or proton radiography data corresponding to plasma `jets' that are attributed to the release of magnetic energy in the form of plasma kinetic energy. These experiments have been focused on observing the outcome of a reconnection `event', but a key missing ingredient in these investigations is the mechanism by which the interaction occurs in the first place. It is this mechanism we are able to observe through lack of field pile-up \cite{Fox2011} and the redistribution of fields creating distinct signatures in the proton probing data.  
 \begin{figure}
 	\centering
	\subcaptionbox{\label{fig:fosetup}}{\includegraphics[width=0.4\textwidth]{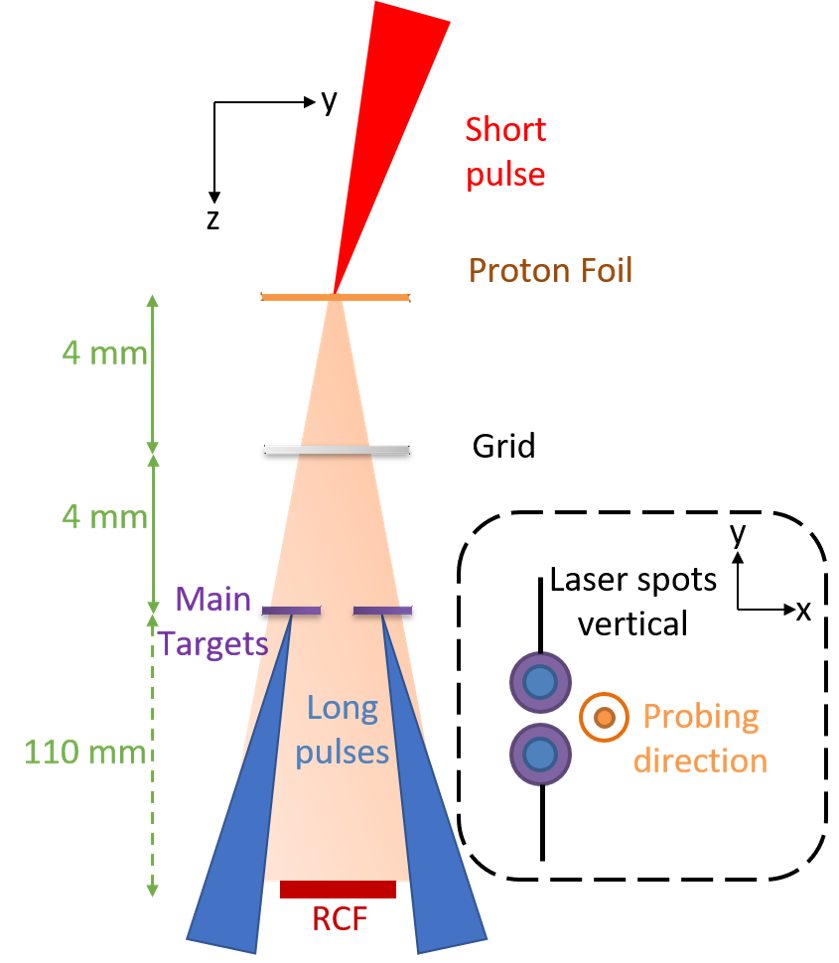}}
	\subcaptionbox{\label{fig:pulseshape}}{\includegraphics[width=0.5\textwidth]{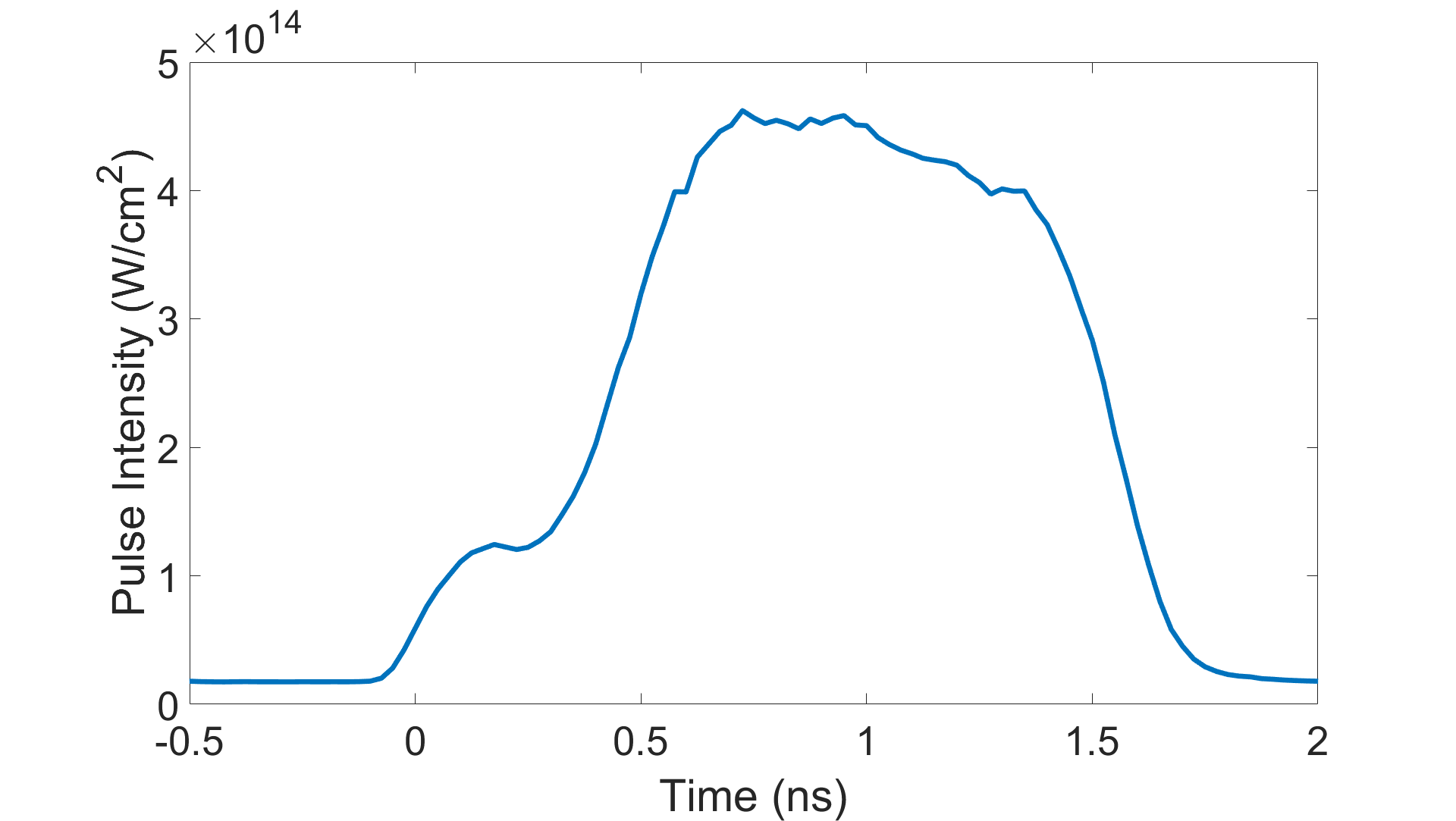}}
	\caption{(a) A diagram showing the orientation of the main target where two laser beams each of 400 J in a \mbox{1.5 ns} stepped pulse (shown in (b)) are focussed onto 400 $\mu$m diameter CHCl discs held by carbon fibres on to an F-shaped mount. The discs were separated by 800 $\mu$m from centre to centre.} \label{fig:setup}
\end{figure}

We determine that in semi-collisional environments the interaction region is governed by the magnetic-field-carrying electrons with a long mean-free-path. The laser pulse interaction with a solid target causes an expanding plasma bubble, which carries with it a frozen-in magnetic field formed by the Biermann battery mechanism \cite{Stamper1971}. These fields are orientated parallel to the target, such that in the centre between two laser spots the fields are oppositely directed. When the two expanding plasmas collide in this central region, the pressure increase causes the plasma flows to slow down and stall. Opposing magnetic fields still frozen to the flow prevent the interpenetration of the two plasma bubbles. Eventually these fields decouple from the flows, allowing rearrangement of the field geometry.An anisotropy in the electron velocity distribution allows more plasma flow out between the bubbles in the x-direction parallel to the magnetic fields, ensuring the fields do not continue to pile up and grow in magnitude.

Crucially, the electron pressure tensor term in Ohm's law reflects this anisotropy and becomes the main support of an electric field within the central region. This effect mediates the change in the magnetic field structure ($\frac{\partial \mathbf{B}}{\partial t} =- \nabla \times \mathbf{E}$) \cite{Hesse1999,Divin2010}, rather than reorganisation from classical resistivity effects. The results to be presented here demonstrate this by showing that the experimental results cannot be replicated quantitatively without considering the contribution from anisotropic pressure.

Here we report on experimental observations, collected at the Orion laser facility, Aldermaston (UK) \cite{Hopps2015}, of magnetized plasma interactions where anisotropic pressure effects are crucial in interpreting the measurements. A series of time-resolved proton radiographs \cite{Mackinnon2004} help to understand the importance of various physical effects that influence the dynamics occurring within the interaction region, typically the potential site for reconnection or diffusion to occur within. The experimental results are supported by kinetic simulations and reconstructed magnetic field maps.

\section*{Results}
\textbf{Proton probing in multiple directions of magnetic field structures.} Two separate laser beams ($\lambda=351$ nm) of 400 J in a 1.5 ns pulse with a temporal profile shown in figure \ref{fig:pulseshape} were focused to a peak intensity of $4.5\times 10^{14} ~\mathrm{W/cm^2}$. The beams were incident at 27$\degree$ to the target normal onto two individual 400 $\mu$m diameter plastic disc targets of 25 $\mu$m thickness. Phase plates were used to smooth the intensity profile in each focal spot, creating elliptical spots of 220 $\mu$m $\times$ 150 $\mu$m in diameter. The ellipse was orientated such that the major axis of each spot was aligned horizontal and parallel to the other. The plasma and fields generated were primarily diagnosed using proton radiography. The probing protons were produced via the target normal sheath acceleration mechanism \cite{Snavely2000,Wilks2001,Borghesi2006} from a $\lambda=1053$ nm wavelength, short pulse beam, focused onto a \mbox{25 $\upmu$m} thick Au target with average intensity, $I=1\times 10^{20} ~ \mathrm{W/cm^2}$. Radiochromic film (RCF) was positioned $110~\textrm{mm}$ from the main interaction target, producing $\times14.75$ magnification of the plasma at the film. The set-up of the experiment is shown in Fig. \ref{fig:fosetup}, in the `face-on' probing arrangement. In this geometry the protons are primarily deflected by magnetic fields orientated perpendicular to the probing direction. Electric fields in this orientation are predominantly in the same direction as the proton probing axis and therefore will minimally affect the proton trajectories.

\begin{figure*}[h]
	\centering
	\parbox{\textwidth}{
		\centering
		\begin{subfigure}{0.35\textwidth}
			\includegraphics[width=\textwidth]{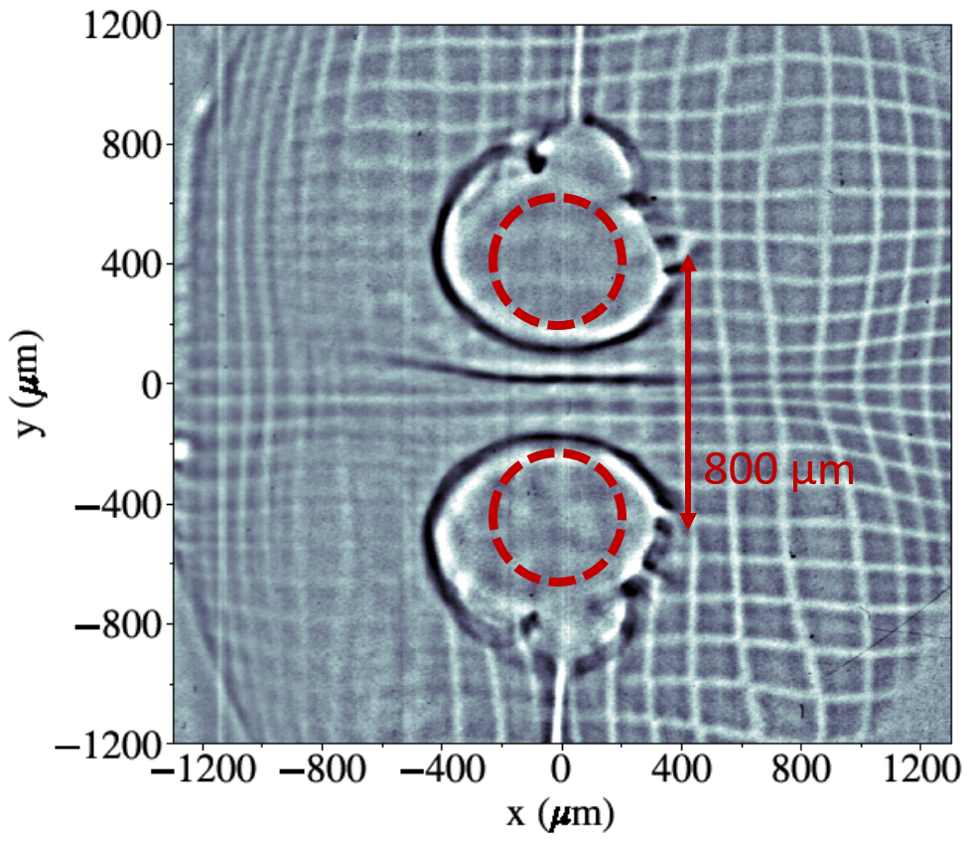}
			\caption{}
			\label{fig:02ns-exp}
		\end{subfigure}
		\begin{subfigure}{0.35\textwidth}
			\includegraphics[width=\textwidth]{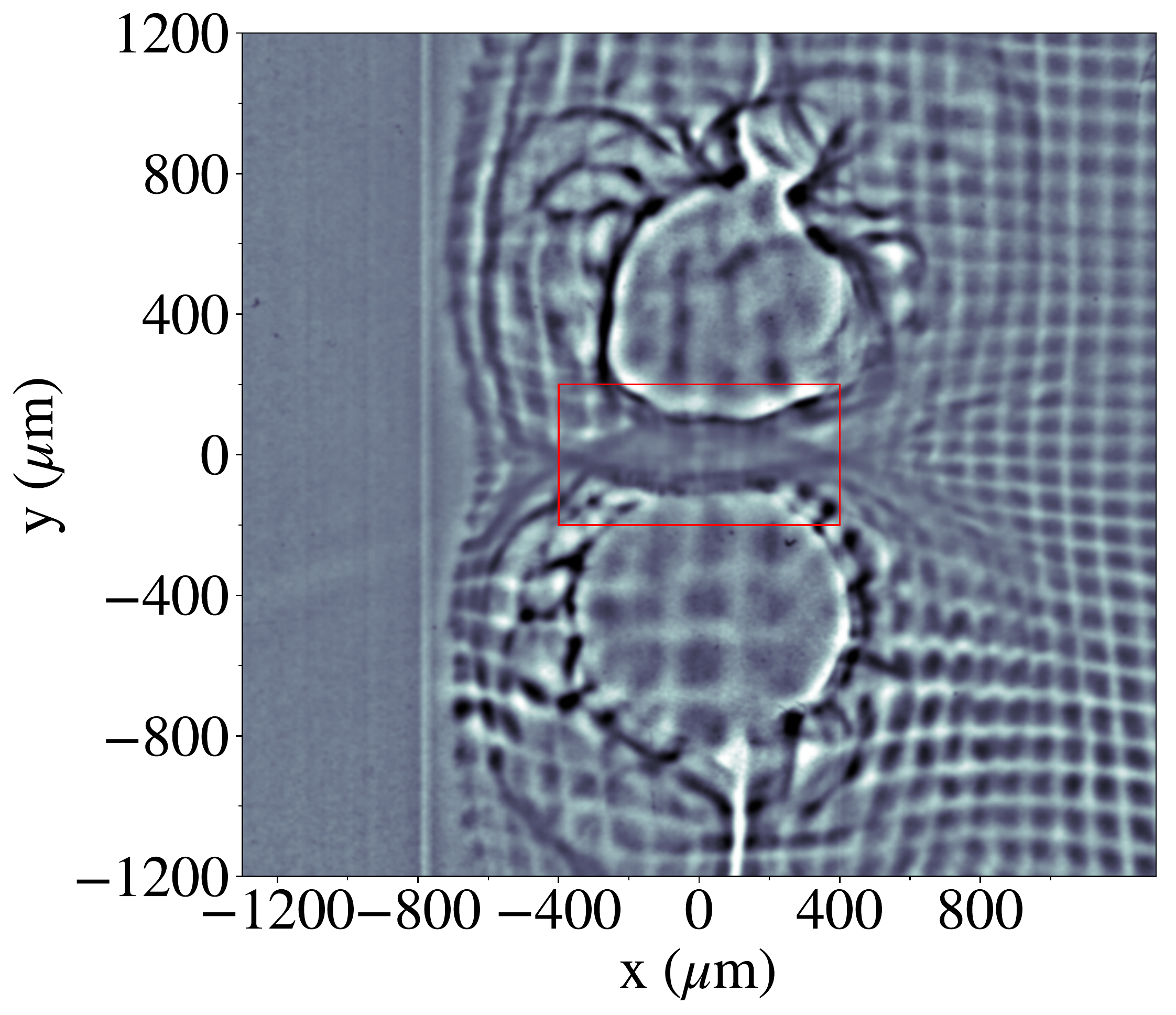}
			\caption{}
			\label{fig:05ns-exp}
		\end{subfigure}
		\\
		\begin{subfigure}[b]{0.35\textwidth}
			\includegraphics[width=\textwidth]{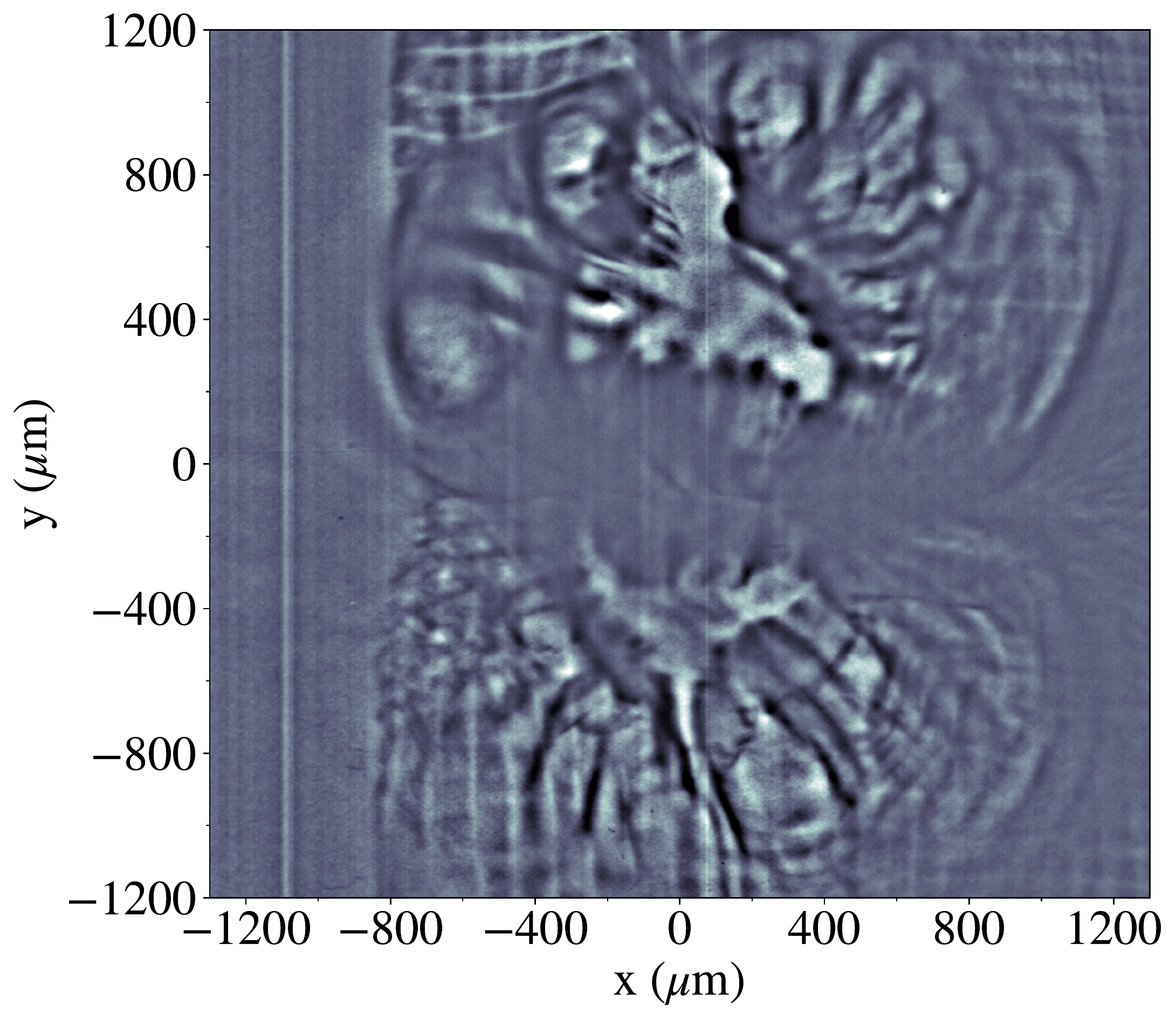}
			\caption{} 
			\label{fig:1ns-exp}
		\end{subfigure}
		\begin{subfigure}[b]{0.35\textwidth}
			\includegraphics[width=\textwidth]{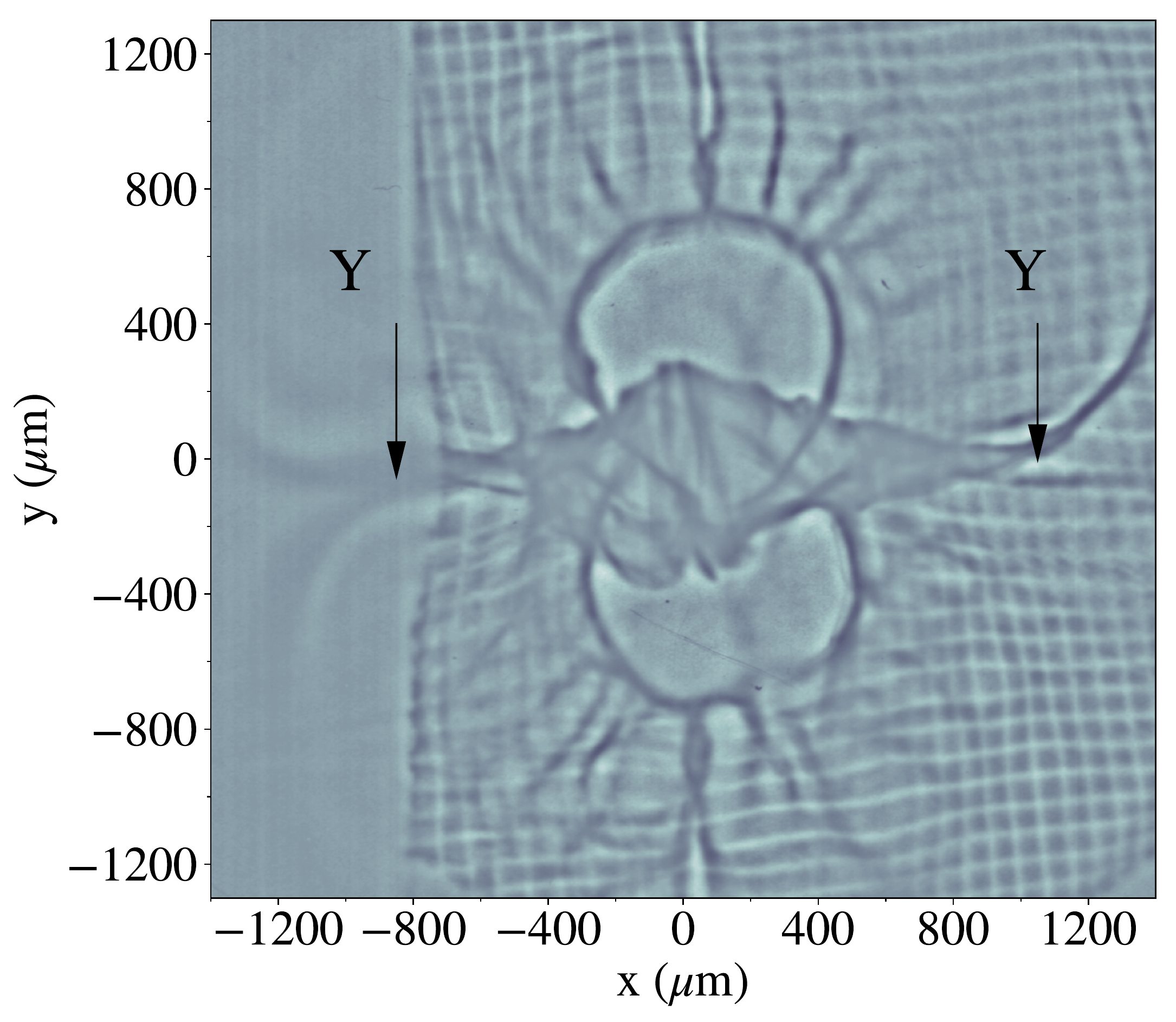}
			\caption{} 
			\label{fig:15ns-exp}
		\end{subfigure}
	}
	\caption{The raw proton radiographs recorded at (a) t = 0.2 ns, (b) 0.5 ns, (c) t = 1.0 ns and (d) t = 1.5 ns. 17.4 MeV protons produce the radiographs shown in (a), (c) and (d) and 15.6 MeV protons produce (b). The image contrast has been adjusted to enhance the features in the radiographs. The red circles in (a) represent the approximate position of the original target discs. The points labelled `Y' in (d) represent the region the bubbles start to separate away from each other.}
	\label{fig:radiographs}
\end{figure*}

Key proton radiographs, recorded at specific times during the long-pulse interaction time, are shown in Figure \ref{fig:radiographs}. At early-time, Figure \ref{fig:02ns-exp}, the evolution of the Biermann battery generated magnetic fields around the laser spots is recorded. The protons are deflected radially outwards by the fields and dark outlines of rings can be observed, particularly at later probing times, surrounding the laser spot region and additional, larger rings formed by the magnetic fields that are being generated and transported by the expanding plasma \cite{Gao2015}. 

It is observed by 0.5 ns (Fig. \ref{fig:05ns-exp}) that the two expanding plasma bubbles mapped out by the protons have overlapping circle outlines within the central region. The overlapping of the two `ring' features around the laser spots observed in the protons does not indicate the two plasmas have necessarily collided at the target. These features can also be caused by proton trajectories that pass through strong fields at the interaction causing large angle deflections resulting in crossing of the protons' paths behind the target. 

By 1 ns (Fig. \ref{fig:1ns-exp}) the plasma and fields have further advected radially outwards with a flow velocity of \mbox{800-1000 km/s} and now have collided. This is evident from the uneven distribution of protons in between the two spots. Using the techniques described in the methods section measurements of the path integrated magnetic fields are extracted. At 1 ns the magnetic fields have a strength of \mbox{50$\pm$5 T}, assuming a magnetic field structure with out-of-plane length, $dl$, of \mbox{350$\pm 25$ $\mu$m}. This scale-length for the out-of-plane magnetic fields is taken from 2D hydrodynamic simulations and is a typical value for experiments of this setup \cite{Gao2015,Rosenberg2015,Li2007}.  It is by this time that the expanding plasma bubbles have collided and simulations show the fields are interacting and reorganising in this region. At the edges of the interaction region, where the two plasma bubbles start to separate away from each other (we call these `Y' regions, as labelled on \ref{fig:15ns-exp}), we observe enhanced darkening around the edges of the spots. The protons are deflected out of the bubbles by smaller amounts, suggesting that towards the edges, away from the central interaction region, the magnetic fields are weakening and being dissipated.

At 1.5 ns (Fig. \ref{fig:15ns-exp}) the magnetic field is \mbox{55$\pm$5 T}, similar to earlier times, due to a pressure anisotropy developing in the electron distribution. Rather than the plasma stagnating and building at the centre the plasma is instead redirected out in the x-direction (horizontal axis in figs. \ref{fig:05ns-exp}-\ref{fig:15ns-exp}). The plasma flow velocity is still high at \mbox{$\sim$800 km/s}, $\sim13 v_A $ (taking $n_i=2.2\times10^{19}~\mathrm{cm^{-3}}$ calculated from simulation), suggesting that the overall plasma bubble expansion velocity is near constant in the unimpeded direction away from the central collision region over the duration of the laser pulse. In comparison, the speed with which the central interaction region expands in the y direction is negligible, and is therefore not governed by the rate of inflow plasma.

\begin{figure*}
	\parbox{\textwidth}{
		\begin{subfigure}{0.30\textwidth}
			\includegraphics[width=\textwidth]{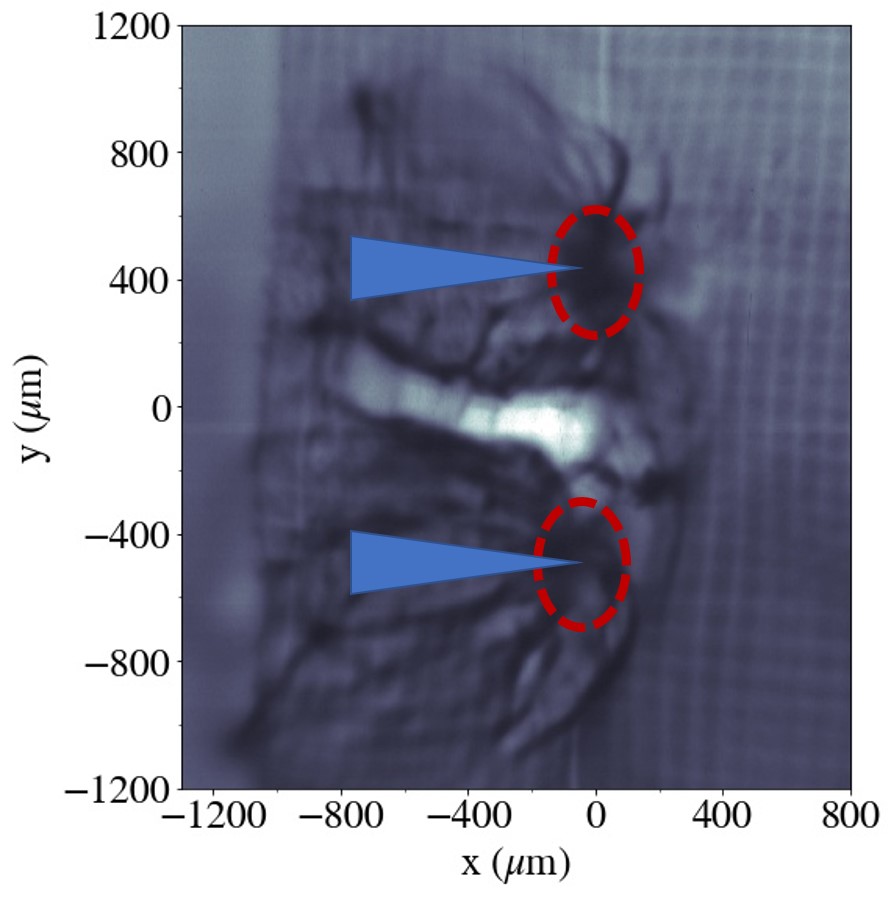}
			\caption{}
			\label{fig:1nsdiag}
		\end{subfigure}
		\begin{subfigure}{0.30\textwidth}
			\includegraphics[width=\textwidth]{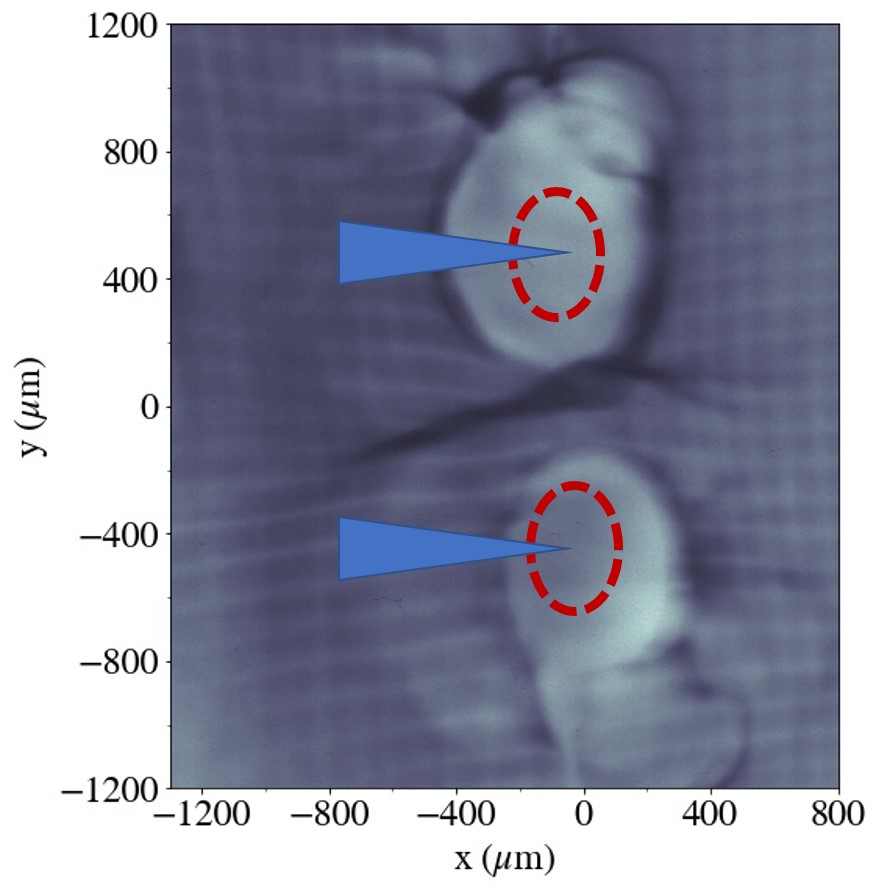}
			\caption{}
			\label{fig:1nsdiagos}
		\end{subfigure}
		\begin{subfigure}{0.30\textwidth}
			\includegraphics[width=\textwidth]{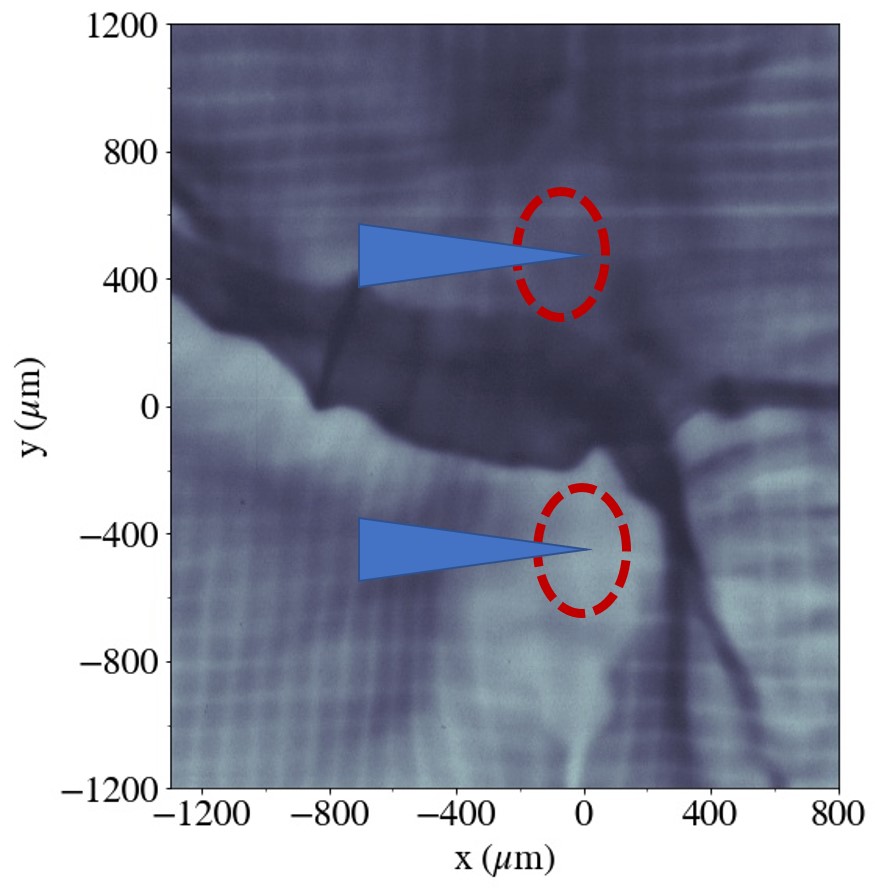}
			\caption{}
			\label{fig:25nsdiag}
		\end{subfigure}
	}
	\caption{Radiographs of the interaction using 17.4 MeV protons probing at 1 ns for (a) and (b) and at 2.5 ns for (c). The protons probe the interaction at 45$\degree$ to the target normal with the protons probing through from the same side as the main laser (a) and from the opposite side (b) and (c). In these images (blue) lasers are incident from the left onto the targets, noted by red ovals.  \label{fig:diag}}
\end{figure*}

In addition to probing through the plasma `face-on' to primarily observe magnetic field deflections we also probed the interaction at 45$\degree$ where protons are more sensitive to both electric and magnetic fields. However by changing the direction the protons probe through the main interaction it is possible to infer the separate influence of these fields on the radiographs. Comparing results from proton probes passing through an interaction from opposite sides will show a reversal in the deflection direction of the protons if the magnetic fields dominate. However if electric fields cause the deflections there will be no change in the deflection direction. The data in Fig. \ref{fig:diag} confirm that deflections in the central region between the laser focal spots are mostly caused by magnetic fields, resulting in dark regions (Fig. \ref{fig:1nsdiagos} and Fig. \ref{fig:25nsdiag}) or light regions (Fig. \ref{fig:1nsdiag}) when the protons come from the same side or the opposite side to the main lasers, respectively. This data also enables reliable estimates of the magnetic and electric fields based on the distortion of a grid placed into the proton beam prior to the main interaction target. In addition to these distortions different energy radiographs can help extract the magnetic and electric fields from the dependence of deflection distance on  proton energy. Magnetic fields measurements are supported using a second method of analysis taking the width of the central regions in a similar technique used for analysis of `face-on' radiographs to calculate the magnetic field. The plasma environment is still fairly complex, however using measured electric field strengths of \mbox{$\sim10^8$ V/m} we can infer reliable measurements of magnetic fields. At 1 ns we estimate a magnetic field of 35$\pm$10 T from the same side probing and 40$\pm$10 T from probing at the opposite side. By 2.5 ns we estimate a field strength of 60$\pm$10 T. 

In figures \ref{fig:1nsdiag} and \ref{fig:25nsdiag} we are also able to see dark lines in the central layer, similar to those we see in Figure \ref{fig:15ns-exp}. These are likely to be due to filamentation instability although their precise origin is a matter of ongoing study as they could be occurring further out from the target surface. \\

\textbf{Reconstruction of magnetic field map from experimental radiographs.} We are able to corroborate the calculations of the magnetic field magnitude measured from proton radiographs using a reconstruction technique that is outlined in ~\cite{Bott2017} and briefly explained later in the methods section. This reconstruction algorithm allows the path-integrated magnetic field to be directly extracted from the proton flux distribution. Provided the proton distribution does not intersect with itself prior to reaching the detector, i.e. only using early time radiographs when the magnetic field gradients are still small, this reconstruction is a mathematically well-defined problem. 

\begin{figure*}[h]
	\centering
	\parbox{\textwidth}{
		\centering
		\begin{subfigure}{0.35\textwidth}
			\includegraphics[width=\textwidth]{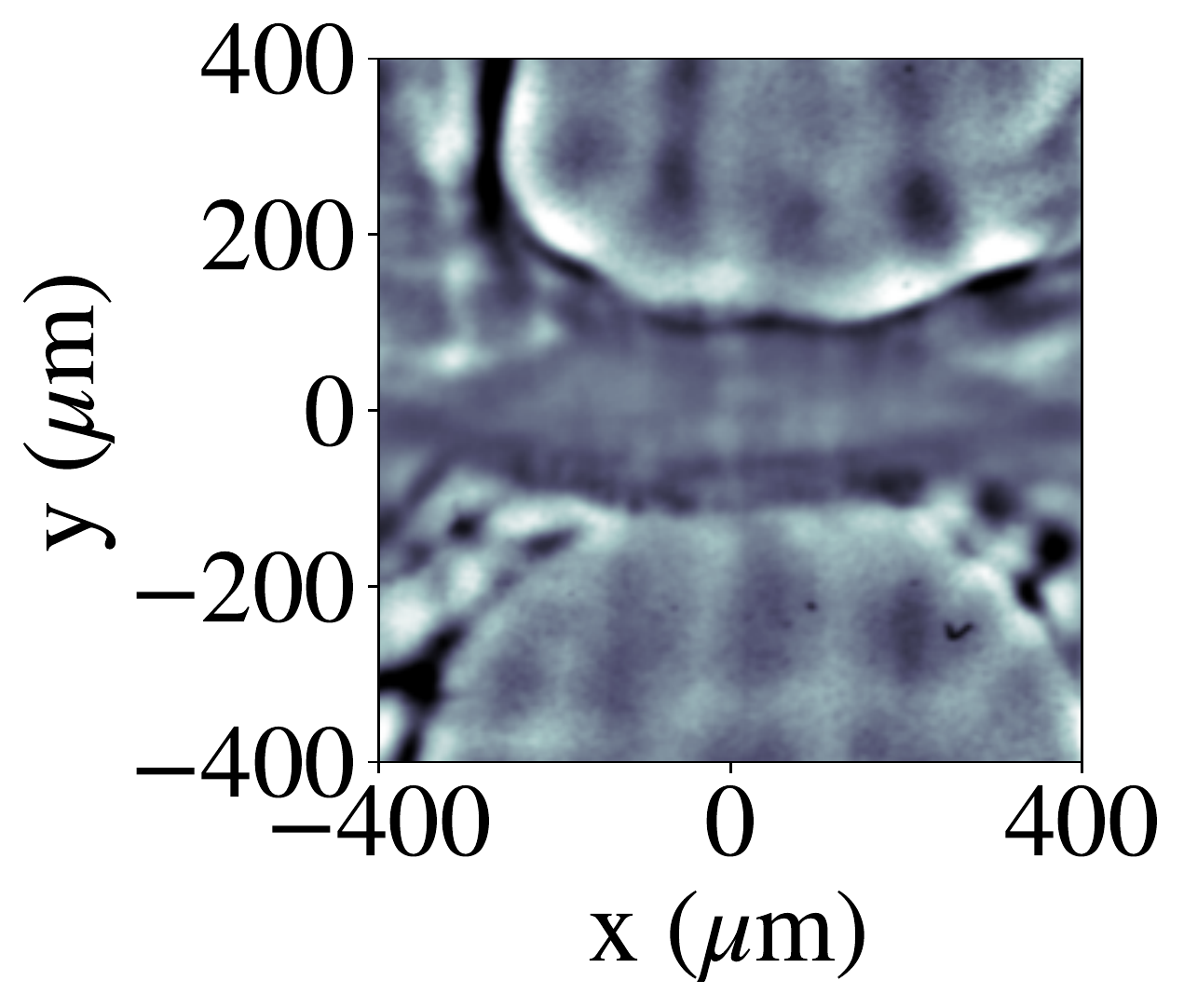}
			\caption{}
			\label{fig:05data}
		\end{subfigure}
		\begin{subfigure}{0.4\textwidth}
			\includegraphics[width=\textwidth]{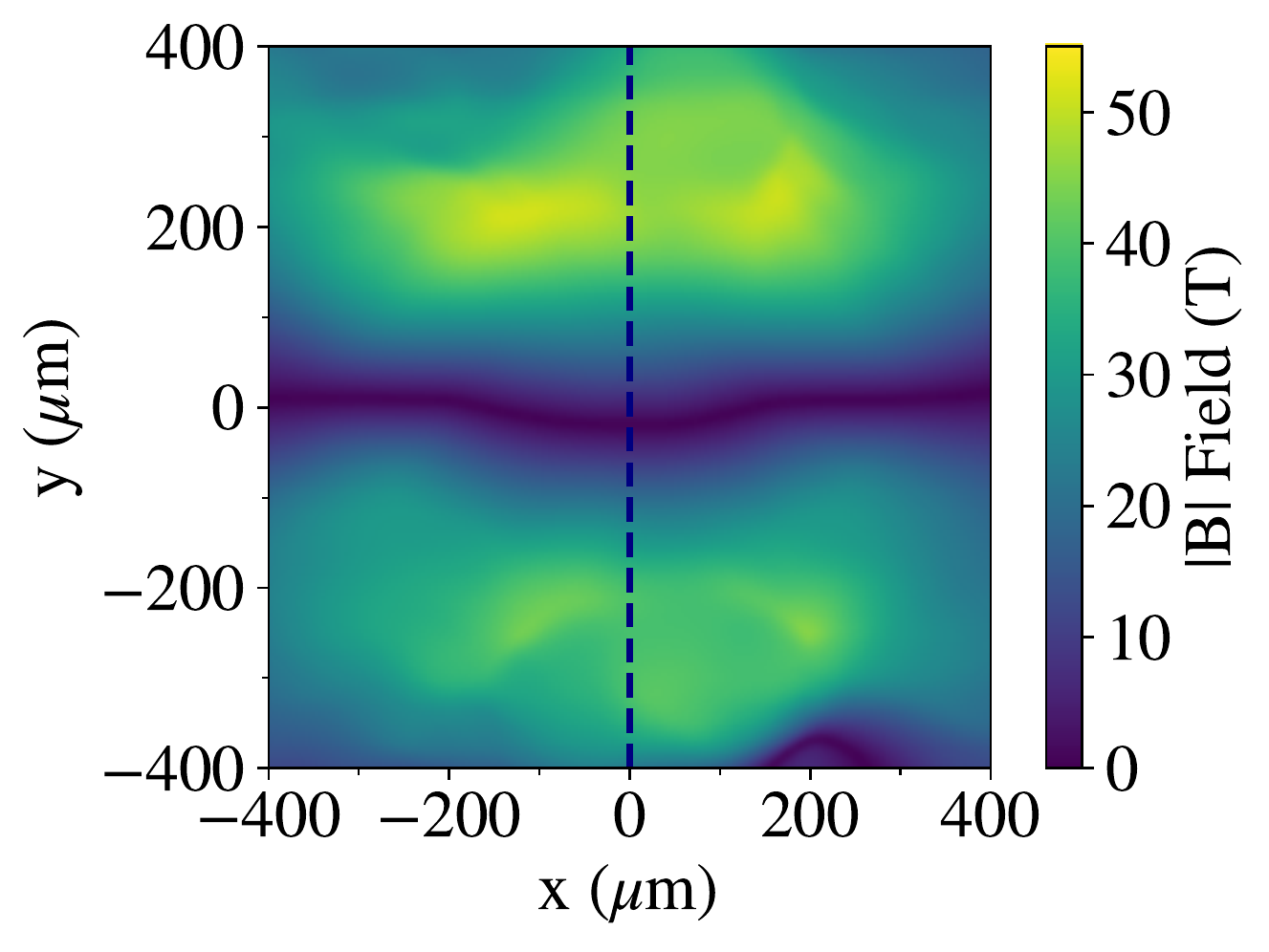}
			\caption{}
			\label{fig:05integrated}
		\end{subfigure}
		\\
		\begin{subfigure}[b]{0.4\textwidth}
			\includegraphics[width=\textwidth]{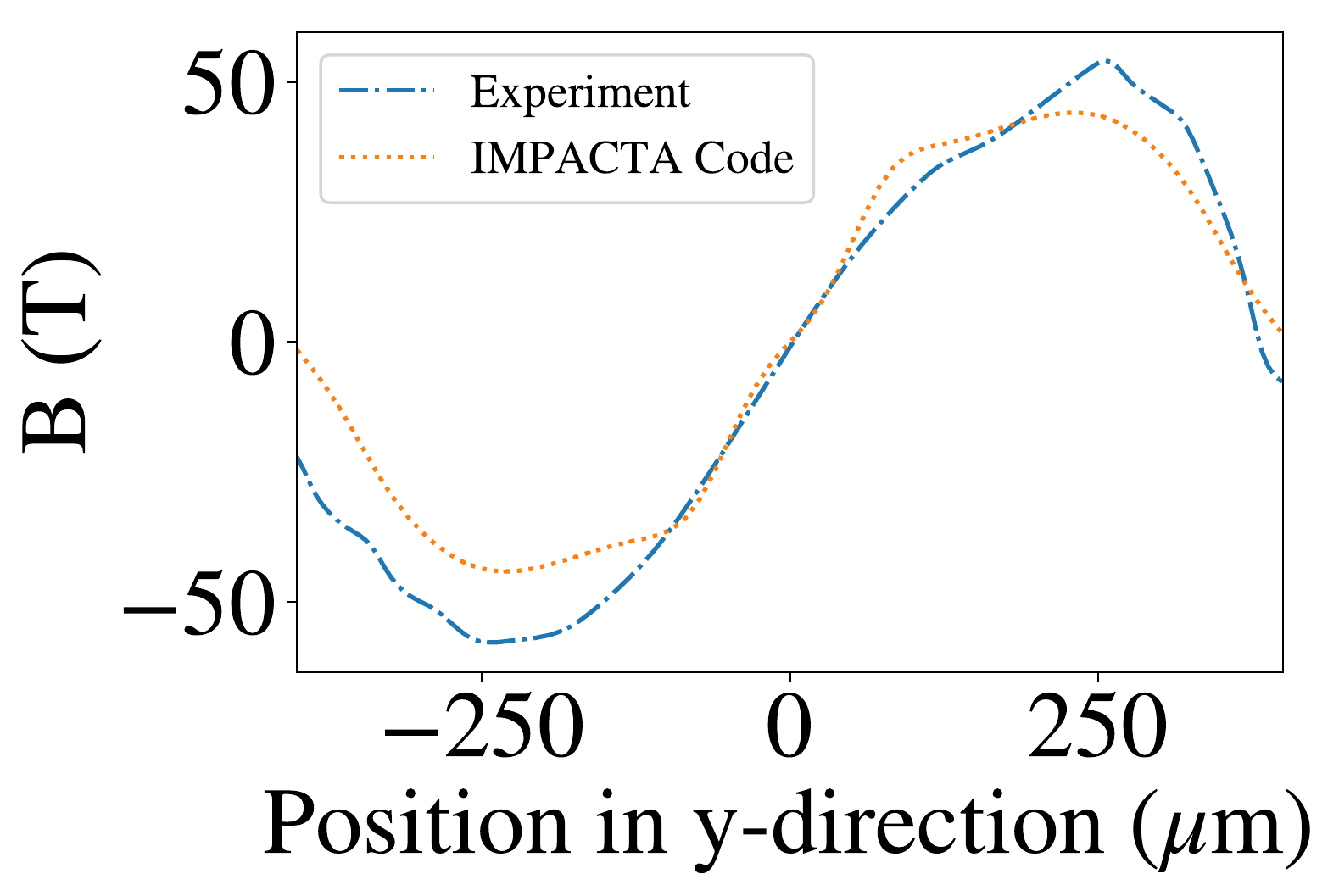}
			\caption{} 
			\label{fig:lineoutintegrated}
		\end{subfigure}
		\begin{subfigure}[b]{0.4\textwidth}
			\includegraphics[width=\textwidth]{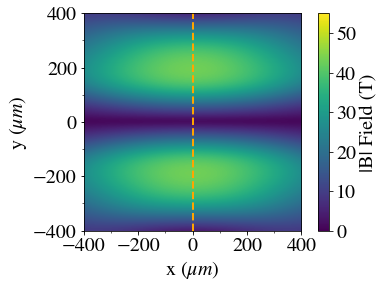}
			\caption{} 
			\label{fig:simucode}
		\end{subfigure}
	}
	\caption{The proton radiograph at 0.5 ns is used to produce a reconstructed 2D map of path-integrated magnetic fields ($B$) using analytical methods \cite{Bott2017}. The whole radiograph is analysed to correctly extract the deflections, although we only show the central region of interest from the experimental radiograph (a) and reconstruction (b) here. The magnetic field strength ($|B|$) along the inflow direction (y) is plotted in (c) from both the reconstructed 2D map and IMPACTA simulations (d), as described in the methods section.}
	\label{fig:pathlengthsflux}
\end{figure*}

Figure \ref{fig:05data} shows an expanded image of the region highlighted by the red box in \ref{fig:05ns-exp}. Using this we create a map of the reconstructed magnetic field, as shown in Figure \ref{fig:05integrated}. Figure \ref{fig:lineoutintegrated} is a lineout of the magnetic field taken in the y-direction at $x=0$ from the reconstructed map and compared to a lineout from magnetic fields produced in kinetic simulations from \textsc{impacta} \cite{Kingham2004,Thomas2009} (Figure \ref{fig:simucode}), which will be described in the following sections. The common form and similar magnitudes of the field recorded in the data and simulations suggests that the simulations do include the correct physics models to match conditions occurring in the experiment. The experimental data shows a peak magnetic field of $\sim$ 50 T at the edge of the laser spots ($|y|=240 ~\mu$m), using the estimated magnetic field structure height of $dl = 200 ~ \mu$m. The magnitude of the \textsc{impacta} simulations, however, at early times slightly underestimate the fields, due to flow velocities being lower than in the experiment (as measured from the expanding plasma bubbles over time), bringing less field into the central region.  \\

\textbf{Using a generalized Ohm's Law to describe plasma dynamics.} In matching the form of the fields inferred from experimental measurement and the field reconstruction analysis, we find the kinetic simulation code must include anisotropic effects. However anisotropy is usually neglected under similar plasma conditions. If neglected this leads to an over-estimate of the magnetic fields piling up between the two plasma bubbles, than those observed experimentally. The role of anisotropic pressure applicable to the conditions created in this experiment is understood through the generalised Ohm's law \cite{Fox2012}, and reproduced here as
\begin{equation}
\mathbf{E}=\bar{\eta}\mathbf{j}+\frac{\mathbf{j}\times \mathbf{B}}{e n_e}-\frac{\nabla \cdot \underline{\underline{P_e}}}{en_e}-\mathbf{v_N}\times \mathbf{B}-\mathbf{v_F} \times \mathbf{B}... \label{eq:ohmslaw},
\end{equation}
where $\eta$ is the resistivity, $\mathbf{j}$ is the current, $\mathbf{B}$ is the magnetic field, $n_e$ is the electron density and $e$ is the electron charge. $\mathbf{v_N}$ is the Nernst velocity \cite{Haines1986} and  $\mathbf{v_F}$ is the plasma flow velocity.  Here, $\eta \mathbf{j}$ is the contribution of the resistive current sheet, ${\mathbf{j}\times \mathbf{B}}/{e n_e}$ is the Hall effect, ${\nabla \cdot \underline{\underline{P_e}}}/{en_e}$ represents the influence of the electron pressure gradients. The $\mathbf{v_N}\times \mathbf{B}$ term describes a bulk field advection term with the electron heat flow and the final term, $\mathbf{v_F} \times \mathbf{B}$, represents magnetic field advection by the plasma flow. Since the flow velocities are much reduced near the region where the two plasmas collide, the two crucial terms that do not diminish in this region are those describing resistivity and electron anisotropic pressure.

The contribution of these two important terms to the electric field is calculated and compared using scalings from ref. \cite{Hesse1999} given by
\begin{align}
\frac{\nabla \cdot \underline{\underline{P_e}}}{e~ n_e}  &\approx \frac{\nabla_y P_\text{yz}}{e~ n_e}  \nonumber \\
       &\approx \frac{1}{e~ n_e}\frac{P_\text{yz}} {\lambda_\text{yz}}  \nonumber \\
       &\approx \frac{1}{e~ n_e ~ \lambda_\text{yz}}    ~ \left(  \frac{ p_e} {2\Omega_e }\frac{\partial v_B}{\partial y}\right)  \nonumber \\ 
       &\approx \frac{m_e v_{th}^2}{4 ~e~ \Omega_e}\frac{v_B}{\Delta_y ~ \lambda_\text{yz}} ,  \label{eq:reduceddivP}
\end{align}
where $\underline{\underline{P_e}}$, $n_e$, $\Omega_e$, $v_{th}$, $m_e$, $\Delta_y$ and  $\lambda_\text{yz}$ are the electron pressure tensor, electron density, electron cyclotron frequency, electron mass, scale length (taken to be the reconnection layer width) and meandering orbit of magnetized electron, respectively. The latter of these is approximated by $\lambda_\text{yz} = \sqrt{\lambda_\text{mfp} r_L}$ where $r_L$ is the Larmor radius and $\lambda_\text{mfp}$ is the collisional mean-free-path of the electron. The values for some of these terms are given in Table \ref{tab:parameters}. Using the plasma conditions in this experiment we are able to approximate the pressure tensor term, $10^5$ V/m $< {\nabla \cdot \underline{\underline{P_e}}}/{e~n_e} < 10^6$ V/m, depending on if one chooses the larger advection velocity from the heat flow, $v_B = 0.4 \kappa \nabla T_e / n_e T_e$ as in ref. \cite{Haines1986,Joglekar2014}, or $v_B = v_F$ where $v_F$ is the flow velocity as in \cite{Rosenberg2015}. 

In comparison, the resistive contribution is small and is approximated by assuming a current established from Ampere's law, $j_z \approx \frac{\partial B_x}{\partial y}\frac{1}{\mu_0} $, which results in currents on the order of $\approx 10^4$. We find $\eta j \approx 10^4$ V/m, a factor of 10-100 smaller than the pressure tensor term, suggesting that the dominant contribution to the electric field in the reconnection region comes from the gradients in the electron pressure tensor. Indeed if temperature gradients are larger and densities are taken to be lower than values currently taken, as might be expected moving even closer to the colliding bubble region, then the anisotropic tensor term will be even more important. 

By estimating the spatial extent of $\nabla \cdot \underline{\underline{P}}_e$, we find further agreement. In this experiment, $n_e \sim 10^{20}$ /cc, and $T_e \approx 1$ keV, measured from experimental diagnostics close to the laser spots and supported by hydrodynamic simulations, giving a typical mean-free-path of $10-50~\mu$m depending on the velocity of the electron. Similarly, the Larmor radius in the presence of a 50 T magnetic field is a few microns but increases near the interaction region where the magnetic field magnitude is small. This gives a meandering orbit \cite{Shuster2015}, $\lambda_\text{yz} \sim 20 ~\mu$m, which suggests that the corresponding electric field created from the electrons is over a length scale comparable to the observed size of the interaction layer, $\Delta_y \approx 50~\mu$m \cite{Ma1996,Liu2017,Ishizawa2004}.

\begin{table*}[h]
	\centering
	\caption{Relevant parameters for calculating contributions to Ohm's Law from experimental data}
	\label{tab:parameters}
	\begin{tabular}{|c|c|c||c|c|c|}
		\hline
		\textbf{Parameter}                           &                             & \textbf{Value} & \textbf{Parameter}                           & \textbf{Scaling}                            & \textbf{Value}   \\ \hline
		$B_0$                               & Observed                   & 10 - 60 T      & $\lambda_\text{mfp}$                & $ v_\text{th} \tau_\text{ei}$  &  $10 - 50~\mu$m           \\ \hline
		$T_0$                               & Observed                   & 1 keV     & $r_L $                               & $ m_e v_\text{th} / e B $                      &   $1 - 6 ~\mu$m                \\ \hline
		$\Delta_y$                     & Observed               & $50 ~\mu$m     & $\lambda_\text{xz}$               & $\sqrt{\lambda_\text{mfp} ~ r_L}$    &   $ 3 - 24 ~\mu$m    \\ \hline
		$\mathbf{\eta j}$                   & $\sim (m / e^2 n_e \tau_\text{ei}) ~ e n_e v_\text{th}$  &   $10^3$ V/m         &
		$\mathbf{\nabla \cdot \underline{\underline{P_e}}{e~ n_e}}$ & eq. \ref{eq:reduceddivP}  & $10^5$ V/m   \\ \hline
		%
	\end{tabular}
\end{table*}

\begin{figure*}[h]
	\parbox{\textwidth}{
		\centering
		\includegraphics[width=0.65\textwidth]{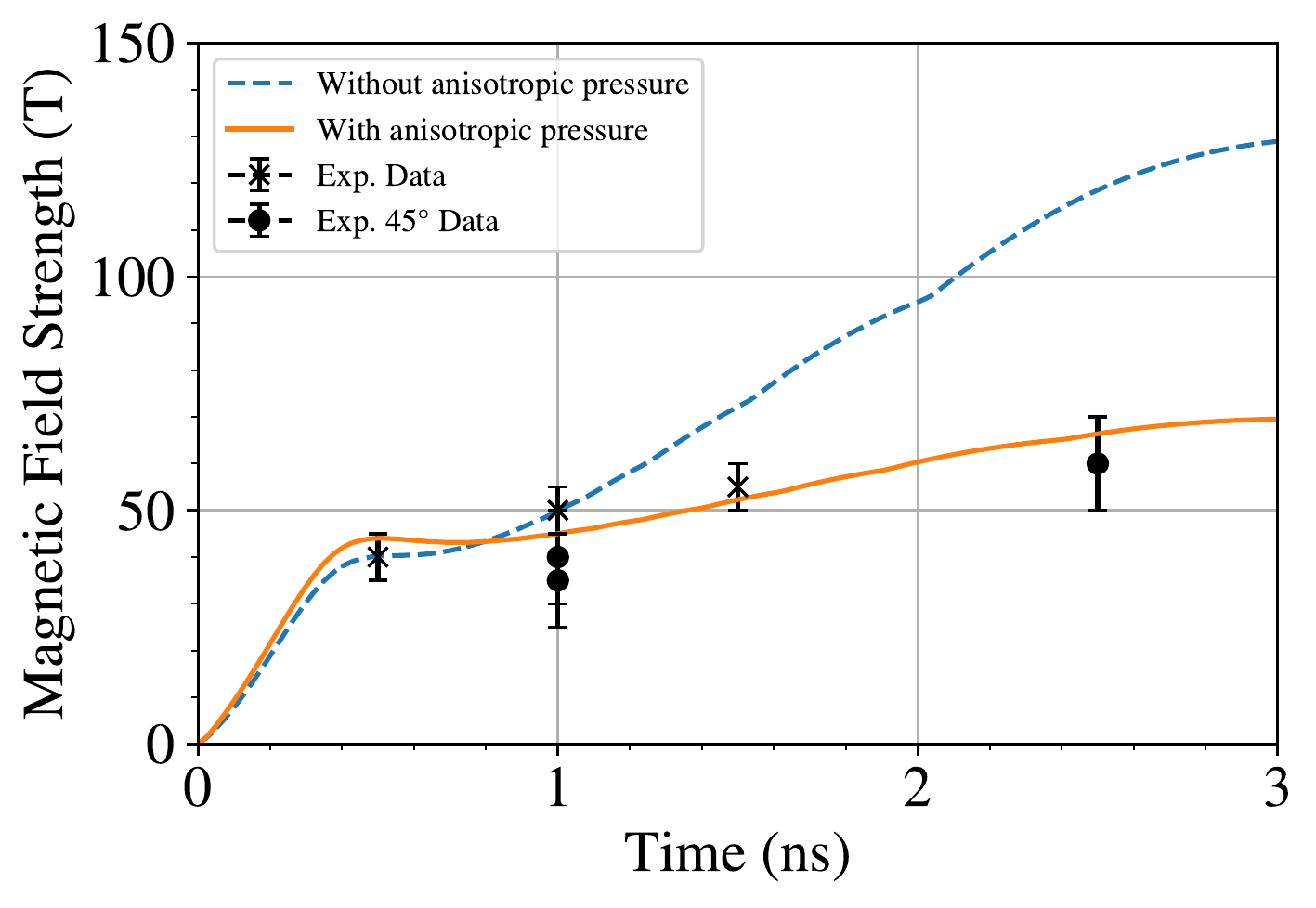}}
	\caption{Numerical modelling of the magnetic field inflow with (orange) and without (blue) the anisotropic pressure term. The resistive approximation results in anomalous magnetic flux-pileup because the electrons are not permitted meandering orbits in the reconnection layer. The inclusion of a 2$^{nd}$ order anisotropy in the kinetic expansion enables this physical effect and reproduces experimental data.}\label{fig:bvstime}
\end{figure*}

\textbf{Numerical simulations in support of experiment.} To support the heuristic deduction that the electric field from the anisotropic pressure is the governing mechanism here, we perform numerical simulations of these experimental conditions using the kinetic code, \textsc{impacta}, a 2D-3V Vlasov-Fokker-Planck-Maxwell model. By choosing to truncate the expansion of the distribution function to only include an isotropic distribution function, $f_0$, and a first order Cartesian Tensor, $\mathbf{f}_1$, \textsc{impacta} includes all the terms in Ohm's Law (eq. \ref{eq:ohmslaw}) except for the anisotropic pressure contribution ($\nabla \cdot \underline{\underline{P}}_e/e n_e$), and can effectively reproduce a kinetic form of resistive MHD by ignoring particular terms. We also run extended simulations using \textsc{impacta} to include the second order Cartesian tensor, $\underline{\underline{\mathbf{f}}}_2$. The magnetic field strengths over time are extracted from simulations including just $\mathbf{f} = f_0 + \mathbf{f}_1 \cdot \mathbf{v}$ and extended to $\mathbf{f} = f_0 + \mathbf{f}_1 \cdot \mathbf{v} +  \underline{\underline{\mathbf{f}}}_2 : \mathbf{v}\mathbf{v}$, i.e. without and with anisotropic pressure effects included respectively, are shown in figure \ref{fig:bvstime}.

In the \textsc{impacta} modelling we find that the magnetic fields collide at t = 0.3 ns, after which time some flux pile-up occurs in both simulations such that the magnetic field value increases by 25$\%$. The two simulations diverge at this point because the simulation that does not include anisotropic pressure enables significant flux-pileup and the magnetic field becomes larger than \mbox{100~ T}. Also plotted are the magnetic fields measured at the central region, between the two spots from the experimental radiographs. The results are shown both from protons probing in multiple directions allowing better understanding and measurement of both magnetic and electric fields. The late time, 2.5 ns, radiograph at 45$\degree$ highlights that the magnetic fields have not strengthened to 130 T, as predicted by simulations when neglecting anisotropic effects, supporting the need for these effects to be included and considered. The time series of proton radiographs have been collected using separate shots of similar conditions. As the data set is small it is insufficient to test for shot-to-shot fluctuations, we conducted a series of simulations varying plasma parameters to ensure that our interpretation of trends in the data is robust. We find that reasonable changes in plasma conditions do not significantly affect our conclusions and agree with theory-driven scalings. This gives us confidence in our analysis. We also observe in the simulations that anisotropic pressure effects allows the magnetic fields to weaken at the edges of the colliding bubbles, agreeing with our interpretation of the experimental results. In comparison, \textsc{impacta} calculations without anisotropic effects on the fields show stronger fields over more extended regions from in between the two bubbles, where the flows and associated magnetic fields have stagnated.

This analysis suggests that an Ohm's Law that includes anisotropic pressure is essential towards reproducing the magnetic field measurements over a nanosecond-long time-scale. Resistive MHD allows significant flux-pileup such that the magnetic field reaches magnitudes not supported by any of our experimental data. 

\section*{Discussion}
In this experiment, we observe the collision between two laser-produced magnetized plasma bubbles and measure a relatively constant value of magnetic field strength over a nano-second time-scale in their interaction region. In order to explain  the magnetic field dynamics, we calculate the magnitudes of the two physical effects most likely responsible in governing the electric field in the interaction region. Using experimental data from multiple probing angles as inputs, the resistive electric field, $\eta \mathbf{j}$, is determined to be negligible in comparison to that created due to gradients in the electron pressure tensor, $\nabla \cdot \underline{\underline{\mathbf{P}}}_e / e n_e $. We also determine that the length scale that governs these dynamics, the meandering orbit of a weakly collisional electron gyrating in a weak magnetic field, $\lambda_\text{xz}$, is comparable to the measured size of the interaction region. 

Using this understanding, we turn to kinetic numerical simulations to test our calculations of the contributions of both physical effects and their relevant length scales. We find that when both, resistive and $\nabla \cdot \underline{\underline{P}}_e$, effects are included in the simulation, the magnitude and shape of the magnetic field agrees well with experimental measurements over the course of a nano-second. Meanwhile, simulations that only include resistive effects give magnetic fields which are pinched along the region between the two colliding plasmas and with significantly higher magnitudes than observed in experiment.  This demonstrates the importance of anisotropic pressure terms to be considered in semi-collisional environments, not just those which are collisionless. The pressure anisotropy is an outcome of an anisotropy in the electron velocity distribution. It represents the ability of the electrons to decouple from the magnetic field due to the current sheet, similar to what occurs in a purely resistive scenario. The decoupling results in reduced field pile-up as the electrons are allowed to redistribute the magnetic field. It should also be noted that this is just one example of conditions under which the pressure anisotropy develops. Future investigations could look to drive weaker and stronger pressure anisotropies by modifying initial conditions to further understand the effect of pressure anisotropy in magnetic flux pile-up.

While we can confidently state that we observe the interaction of magnetic fields mediated by anisotropic pressure, we remain hesitant to state that we observe magnetic reconnection mediated by the same because we do not observe direct evidence of this. Partially, this is because we have a lack of plasma `jets' which have been cited as evidence of magnetic reconnection in previous experiments \cite{Nilson2006,Rosenberg2015}. However, we suggest that `jets' are not essential for magnetic reconnection in a collisional plasma because the release of magnetic field energy may translate to heat flow rather than particle acceleration. To determine whether magnetic reconnection has occurred in these semi-collisional, laser-plasmas, experiments must determine inflow and outflow profiles of plasma temperature and magnetic fields. Understanding a power balance of these energies \cite{Hare2017} might then help consider if reconnection is a process occurring. The work presented here suggests that the magnetic field evolution in the interaction region will still be governed by the anisotropic pressure gradients.

\begin{footnotesize}
\subsection*{Methods}

\subsubsection*{Proton Radiography}
Target normal sheath acceleration produces protons of up to 40 MeV from a thin gold foil and subsequently used to probe the fields of a plasma. Radiochromic film is used to record the final position of the protons, and therefore infer the fields causing the deflection from their unperturbed positions. Stacks of radiochromic film are layered with filters of iron to allow for measuring of higher proton energies without the need for large stack dimensions. The resulting radiochromic films are scanned and analysed to extract the magnetic fields by either noting the distortion caused to the grid, or by half the width of the central darker region in between the two spots. The grid is imprinted on the beam before the protons are sent across the main interaction. Without any fields a magnified grid structure would be produced at the RCF and so this position can be compared to the deflected grid. At early times the grid is very visible and this method is more accurate, however at 1 ns the grid is less pronounced and so other methods are used. The second method is to take half the width of the central dark feature, $\Delta y$, assuming that a proton originally at the edge of one plasma bubble is deflected towards the second plasma bubble.  Both the grid and half width methods calculate magnetic fields that agree in magnitude. These methods also support the fields predicted from the path integrated reconstruction results. 

The proton deflection at the RCF is used to calculate the fields using the Lorentz force equation yielding either the magnetic or electric field contribution. The set-up shown in Fig. \ref{fig:fosetup} allows probing of primarily magnetic fields. The magnitudes of these fields are given by:

\begin{equation}
\int B\times dl=\frac{Md}{eb}\sqrt{2m_pE_p}
\label{eq:defle}
\end{equation}

In this equation $d$ is the maximum displacement of the proton from its normal trajectory as recorded at the film, $M$ is the magnification of the target at the RCF, $e$ is the charge of the proton, $b$ is the length the proton travels after the interaction region, $E_p$ is the proton energy and $\int B\times dl$ is the integrated path length of the magnetic fields that the proton travels through \cite{Li2006,Li2007}. The length $dl$ is consistent with both scales in hydrodynamic simulations extracted from using both the HELIOS-CR software package \cite{MacFarlane2006} and the NYM Lagrangian code \cite{Roberts1980}, in a similar manner to previous experiments \cite{Gao2015,Rosenberg2015}. Table \ref{table:Bdl} shows the extracted measurements for each radiograph and the different scale-length values taken for each. These have been verified across the different energy radiographs and closely agree with each other. When considering the 45$\degree$ radiographs the length, $dl$, taken also has to be adjusted as well as the magnetic fields acting at an angle causing deflections. These factors in combination with contributions of electric field deflections means that the errors associated with these measurements are larger, however they still give a good guide and help constrain our understanding of the late time evolution.

\begin{table*}[t]
	\centering
	\caption{Extracted path integrated magnetic fields from experimental proton radiographs.}
	\label{table:Bdl}
	\begin{tabular}{|c|c|c|}
		\hline
		Time (ns)& $dl$ ($\upmu$m)&             $B$ (T) \\    
		 \hline
		 Face-on& &  \\
		 0.5 &  150 $\pm$ 50  & 40 $\pm$ 5\\
		 1 &  350 $\pm$ 50 &  50 $\pm$ 5\\
		 1.5&  450 $\pm$ 50&  55 $\pm$ 5\\
		 \hline
		 45$\degree$ & & \\
		 1 (same side)&350 $\pm$ 50 &35 $\pm$ 10 \\
		 1(opposite side)& 350 $\pm$ 50&40 $\pm$ 10 \\
		 2.5 (opposite side) &600 $\pm$ 50 &60 $\pm$ 10 \\
		 \hline
	\end{tabular}
\end{table*}

\subsubsection*{Path Integrated Field Reconstruction}
The extraction of the path-integrated field from early-time proton radiographs is a multi-step process. First, the image recorded on the Gafchromic EBT3 or HD-V2 RCF film stack must be converted into the proton flux distribution relative to some mean flux. This is done by converting the measured optical density into an estimate of the dose~\cite{Schollmeier2014}. Once this conversion is completed, a spatial filter removing low wavelengths is applied to the proton flux distribution. This is because the reconstruction process is sensitive to large-scale variations in the proton flux distribution which are the result not of deflections by magnetic fields, but by unmeasured variations in the initial TSNA proton flux. However, these variations occur on larger scales than the order-unity relative flux inhomogeneities observed in the experimental images, and so the impact of these variations can be removed with the filter~\cite{Manuel2012}. 

The proton radiography diagnostic used on this experiment reasonably satisfies the paraxial approximation: that is, the distance $r_i = 8$ mm from the proton foil to the target exceeds the perpendicular extent $l_\bot \sim 1$ mm of the main targets. In this case, the proton distribution can be well described as a two-dimensional sheet travelling in a single direction (which we denote $\hat{\mathbf{z}}$ here). Furthermore, provided the angle of deflections of the protons due to magnetic fields are small, it can be shown that the deflection velocity $\delta \mathbf{v}\!\left(\mathbf{x}_{\bot0}\right)$ perpendicular to the $z$-direction of a proton with an initial perpendicular position $\mathbf{x}_{\bot0}$ in the plasma plane is given by
\begin{equation}
\begin{split}
\delta \mathbf{v}\!\left(\mathbf{x}_{\bot0}\right) &\approx \frac{e}{m_p c V} \nabla_{\bot0} \left[\int_0^{l_z} \mathrm{d}z' \; A_z\!\left(\mathbf{x}_{\bot0}\left(1+\frac{z'}{r_i}\right),z'\right) \right] \\
&= \nabla_{\bot0} \varphi\!\left(\mathbf{x}_{\bot0}\right) \, ,
\end{split}
\end{equation}
where $e$ is the elemental charge, $m_p$ the proton mass, $c$ the speed of light, $V$ the initial proton velocity, $l_z$ the parallel extent of the plasma, $\mathbf{A}$ the vector potential for the magnetic field, $\nabla_{\bot0} \equiv \partial/\partial \mathbf{x}_{\bot0}$ a gradient operator with respect to the initial perpendicular plasma coordinates, and $\varphi = \varphi\!\left(\mathbf{x}_{\bot0}\right)$ a scalar function. Finally, if the adjacent regions of the proton beam do not overlap as a result of these deflections, the proton flux distribution $\Psi$ is related to the scalar function $\varphi$ via a Monge-Amp\`ere equation of the form
\begin{equation}
\begin{split}
\Psi\!\left(\nabla_{\bot0} \phi\!\left(\mathbf{x}_{\bot0}\right) \right) = \frac{\Psi_{0}\!\left(\mathbf{x}_{\bot0}\right)}{\det{\nabla_{\bot0} \nabla_{\bot0} \phi\!\left(\mathbf{x}_{\bot0}\right) }} \, , 
\label{MongeAmpere}
\end{split}
\end{equation}
where $\Psi_0$ is the initial proton distribution, and $\phi\!\left(\mathbf{x}_{\bot0}\right) \equiv \left(r_s + r_i\right) \mathbf{x}_{\bot0}^2/{2r_i} + r_s \, \varphi\!\left(\mathbf{x}_{\bot0}\right)/V$, for $r_s$ the distance from the plasma to the screen. The deviation of these results has been described fully elsewhere, as is the numerical inversion procedure for recovering $\phi$ from \eqref{MongeAmpere} given appropriate boundary conditions (that is, assuming no proton is deflected off the RCF stack)~\cite{Bott2017,Sulman2011}. Once $\phi$, and hence $\varphi$, has been determined, this allows for the calculation of the path-integrated $z$-component of the vector potential; taking the curl of this quantity gives the path-integrated perpendicular magnetic field. 

For our particular images, the initial proton flux in regions passing through the main target circular foils were reduced in line with the observed reduction in proton flux seen in those regions. In principle, it might be expected that the presence of proton flux variations due to the presence of a grid might distort the result; however, the inversion procedure for the Monge-Amp\`ere equation is relatively insensitive to periodic variations on smaller scales than the magnetic structures of interest, so it was found that this effect was negligible.

\subsubsection*{Ohm's Law}
Equation \ref{eq:reduceddivP} is derived from the Vlasov-Fokker-Planck equation where the distribution function, $f$, is expanded as a vector, $\mathbf{f}_1$, and tensor perturbation, $\underline{\underline{\mathbf{f}_2}}$, on an isotropic $f_0$ such that $f = f_0 + \mathbf{f}_1 \cdot \mathbf{v} + \underline{\underline{\mathbf{f}}}_2 : \mathbf{vv}+ ...$. The corresponding Vlasov-Fokker-Planck equation for $\mathbf{f}_1$ is 
\begin{equation}
\begin{split}
\frac{\partial \mathbf{f}_1}{\partial t} - v \nabla f_0 + \frac{e\mathbf{E}}{m_e}\frac{\partial f_0}{\partial v} - \frac{e \mathbf{B}}{m_e} \times \mathbf{f}_1 + \frac{2}{5} v \nabla \cdot \underline{\underline{\mathbf{f}_2}} \\= - \frac{Y n_i Z^2}{v^3} \mathbf{f}_1
\end{split}
\end{equation}
Multiplying by $v^6$ and integrating over velocity space gives eq. \ref{eq:reduceddivP}.

\subsubsection*{Modelling}

The kinetic \textsc{impacta} modelling was performed using parameters from early-time, unmagnetized, hydrodynamic simulations. Therefore, the \textsc{impacta} simulation resembled a 2D, initial-value-problem solved over a nanosecond. \textsc{impacta} uses the same expansion as provided in the Methods-Ohm's Law section and solves the coupled set of Vlasov-Fokker-Planck-Maxwell equations for the variables, $f_0, \mathbf{f}_1, \underline{\underline{\mathbf{f}}}_2, \mathbf{E}$, and $\mathbf{B}$ using a fully-implicit method that enables time-steps on the order of the electron-ion collision time. The details are given in refs. \cite{Kingham2004,Thomas2009}. 

Two inverse-bremsstrahlung-heating spots with $I(x,y) = 10^{15}$ W/cm$^2$ were imposed on a uniform density profile. The density was given by the hydrodynamic simulations 0.3 mm from the target surface. An out-of-plane density gradient matching that given by the hydrodynamic simulations was imposed on the system. The heating of the plasma along with an out-of-plane density gradient results in a self-generated magnetic field around the heating regions. The out-of-plane density gradient is relaxed over time to resemble the hydrodynamic simulations, and eventually turned off. During the time it is on, 30 T magnetic fields are generated around each heating region. The magnetic fields are carried towards one another by plasma flow and heat flow.

\subsection*{Data Availability}
The data from the Orion Experiment and the codes for the simulations used in this analysis are available at: \url{https://pure.york.ac.uk/portal/en/datasets/observations-of-pressure-anisotropy-eects-within-semicollisional-magnetizedplasma-bubbles(ed19612c-5e53-4662-b2b9-b46ce72cc09e).html}

\section*{Acknowledgements}
We would like to acknowledge the support and expertise of the Orion technical team at AWE as well as the target fabrication team at the Central Laser Facility.
This work was supported by the United Kingdom Engineering and Physical Sciences
Research Council (grant number EP/K504178/1, EP/J002550/1, EP/L002221/1 and EP/K022415/1) and by the National Science Foundation (grant number 1804463).

\section*{Author Contributions}

The experiment was conceived and designed by N.C.W, E.R.T., M.B., J.M.F., G.G., E.T.G., M.P.H., S.K., R.J.K., C.P.R., A.G.R.T. and L.W. The experimental work at the Orion laser was carried out by E.R.T., A.S.J., B.C., P.D., S.W., T.H. and N.C.W with support from key Orion team members G.C., C.N.D., E.T.G., J.M.F., M.P.H., J.S. and P.T.  Targets were provided by C.S. Experimental data was analysed by E.R.T. The theoretical derivations were developed by A.S.J. with support from M.R., R.J.K., A.G.R.T., E.R.T. and C.P.R. The kinetic simulations were performed by A.S.J and the hydrodynamic simulations were performed by P.G. Reconstruction of experimental data was analysed by A.F.A.B. The manuscript was written by E.R.T, A.S.J. and A.F.A.B with support from N.C.W., L.W. and C.P.R.  

\section*{Competing Interests}
The authors declare no competing interests.

\end{footnotesize}
\end{document}